\documentclass[twocolumn,amsmath,amssymb,floatfix]{revtex4}

\usepackage{graphicx}
\usepackage{dcolumn}
\usepackage{bm}

\setkeys{Gin}{width=\linewidth}

\begin{document}

\title{Rectification in three-terminal graphene junctions}

\author{A. Jacobsen$^1$, I. Shorubalko$^2$, L. Maag$^1$, U. Sennhauser$^2$ and K. Ensslin$^1$}

\affiliation{$^1$Solid State Physics Laboratory, ETH Zurich, 8093 Zurich, Switzerland,\\
$^2$Electronics/Metrology/Reliability Laboratory, Swiss Federal Laboratories for Materials Science and Technology (EMPA), 8600 Duebendorf, Switzerland}


\begin{abstract}

Nonlinear electrical properties of graphene-based three-terminal nanojunctions are presented.
Intrinsic rectification of voltage is observed up to room temperature. The sign and the efficiency
of the rectification can be tuned by a gate. Changing the charge carrier type from holes
to electrons results in a change of the rectification sign. At a bias $< 20$\,mV and at
a temperature below 4.2\,K the sign and the efficiency of the rectification are governed by universal conductance fluctuations.

\end{abstract}

\maketitle


Graphene has attracted a lot of attention due to its unique
properties and a wide field of possible applications. One class of
graphene-based devices which has a realistic chance to reach the
market is ballistic switches and rectifiers that can operate at
ultrahigh-frequencies \cite{geim2009}. The main idea of such devices
is that the charge carriers move without scattering through the
active area of the device or are scattered only by the designed
geometry. Typical examples of such nanodevices are three-terminal
ballistic junctions (TBJs)
\cite{xu2001,worschech2001,shorubalko2001,jordan2008}, ballistic
rectifiers \cite{song1998,song2001}, and artificial nanomaterials
\cite{song2001b}. Three-terminal nanojunctions exhibit
robust nonlinear rectifying behavior in different material systems:
III-V semiconductor heterostructures
\cite{worschech2001,shorubalko2001,shorubalko2003}, InAs nanowires
\cite{suyatin2008} and in carbon nanotubes \cite{papadopoulos2000}.
A number of devices have been demonstrated based on nonlinear
electrical properties of TBJs: frequency multiplier
\cite{shorubalko2002}, logic gates \cite{xu2004,muller2007},
set-reset latch device \cite{sun2008}, half and full adders
\cite{reitzensein2004,lau2006}. The operation speed of such devices
is predicted to be in the THz range \cite{mateos2003}.

Ballistic switches and rectifiers in graphene represent another
fascinating possibility to realize high quality electronic
nanostructures. High charge carrier mobilities are promising for
graphene-based ballistic electronics \cite{bolotin2008} and the possibility to
electrically or environmentally tune graphene from p- to n-type and
vice versa \cite{schedin2007} gives the opportunity to design
adaptive electronic devices \cite{yu2009}. Although theoretical
proposals have been made for building blocks for integrated graphene
circuits \cite{areshikin2007}, to the best of our knowledge none of
the nanodevices mentioned above has been realized in graphene yet.

Here we set out to realize three-terminal nanojunctions in graphene
and investigate their nonlinear electrical properties. Intrinsic
rectification of voltage with a switchable sign is observed up to
room temperature. The possible mechanisms of rectification and
limitations of its efficiency are discussed.


\begin{figure}
  \begin{center}
    \includegraphics{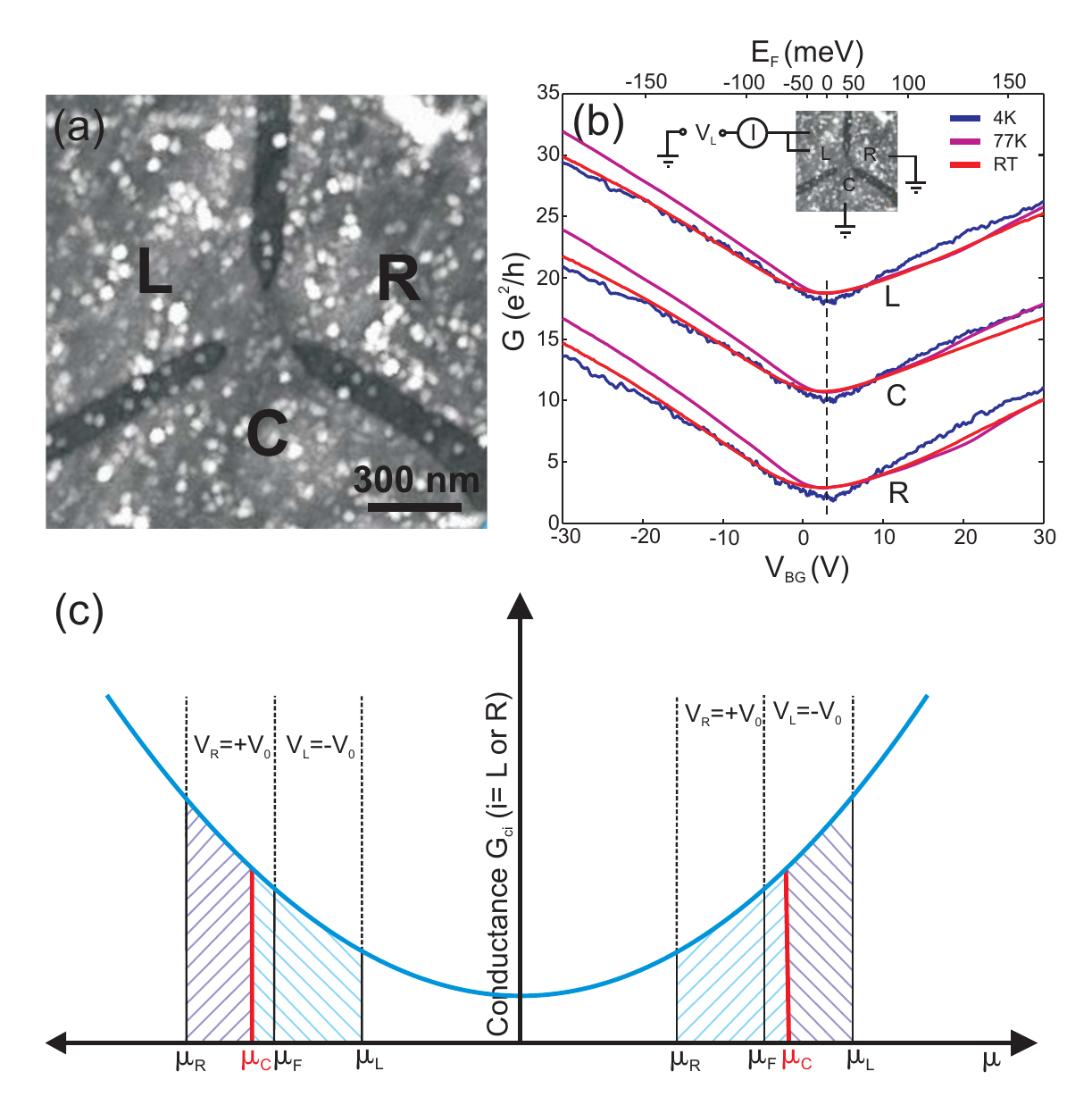}
    \caption{(a) Scanning force micrograph (SFM) of the center part of the
     three terminal junction. The three constrictions, labeled L,R and C, separating
     the center part of the junction from the leads are all $\approx200$\,nm wide. (b) Conductance of each of
     the three constrictions (other two branches grounded) as a function of backgate voltage measured
     at 4\,K, 77\,K and room temperature. The curves of the different constrictions are vertically offset
     by 8$e^2/h$ for clarity. The electrical measurement configuration is shown in the inset.
     (c) A schematic picture which explains the change of sign of the voltage rectification when changing
     from electron transport to hole transport ($V_\mathrm{L}=-V_\mathrm{R}=V_0$). The sign of $V_\mathrm{C}$ is
     independent of the relative signs of $V_\mathrm{L}$ and $V_\mathrm{R}$.}
    \label{fig1}
  \end{center}
\end{figure}

The samples are fabricated by mechanical exfoliation of natural
graphite powder followed by the deposition onto a highly doped
silicon substrate covered by 285\,nm of silicon
dioxide \cite{novoselov2004}. The doped silicon substrate is used as a
global backgate to change the overall Fermi energy of the device.
Single layer flakes are identified by optical microscopy and Raman
spectroscopy \cite{ferrari2006,graf2007}. The Ohmic contacts are
defined by electron beam lithography followed by the evaporation of
Cr/Au (2\,nm/40\,nm), and the structure is patterend in a second electron beam lithography step followed by
reactive ion etching.

In Fig.\,\ref{fig1}(a) a scanning force micrograph (SFM) of the
center part of the device is depicted. The leads, labeled L
(left), R (right) and C (center) are separated by three
constrictions which are $\approx$\,200\,nm wide. The angle between
each branch of the junction is approximately 120 degrees which makes
the device rotationally symmetric and it can be measured in three
comparable configurations in the non-linear transport regime (see measurement configuration below). To
characterize the junction the conductance ($G$) of each constriction
is measured as a function of the applied backgate voltage ($V_{\mathrm{BG}}$) at 4\,K, 77\,K
and room temperature in the linear response regime as shown in Fig.\,\ref{fig1}(b) (The electrical measurement configuration can
be seen in the inset). The Fermi energy calculated from the backgate voltage $V_{\mathrm{BG}}$ by assuming a density of states linear in energy is plotted
as the upper horizontal axis. The point of minimum conductance,
the Dirac point, is located at approximately +3V in backgate (see
the dashed line in Fig.\,\ref{fig1}(b)) and we set the Fermi energy to
be zero at this point. To the left of the Dirac point the charge
carriers are holes and to the right we have electron transport.
Thus, by changing $V_{\mathrm{BG}}$ we can tune the device from
p-type to n-type. At 4\,K reproducible universal conductance
fluctuations (UCFs) are visible. At 77\,K and room temperature the UCFs are smeared out
and the conductance curves are smooth. The conductance of each of
the constrictions is similar suggesting that the device is
close to symmetric. From the conductance curves we estimate the
mobility of the junction to be $\approx 4000$ cm$^2$/Vs and the corresponding mean free path to be $\approx 70$\,nm.
Thus, the device can be considered to be diffusive.


In previous measurements and calculations it has been shown that for
a symmetric III-V-based three-terminal ballistic junction in
the non-linear regime the voltage on the center
branch ($V_\mathrm{C}$) is always negative when the bias voltages at the left and the
right branches are applied in a push-pull fashion with
$V_\mathrm{L}=-V_\mathrm{R}=V_0 $\cite{shorubalko2001,worschech2001,xu2001}. For a
graphene three terminal junction we expect the voltage on the center
branch to always be negative for electron transport and always be
positive for hole transport, independent of the relative signs of $V_\mathrm{L}$ and $V_\mathrm{R}$. This is schematically explained in
Fig.\,\ref{fig1}(c). The conductance from the left branch to the
central branch and the conductance from the central branch to the
right branch increase with increasing chemical potential $\mu$. When
a push-pull bias is applied the chemical potential of the right
branch is $\mu_\mathrm{R}=\mu_\mathrm{F}-eV_0$ and the chemical
potential of the left branch is
$\mu_\mathrm{L}=\mu_\mathrm{F}+eV_0$. For electron transport the area $A_\mathrm{L}$ under
the conductance curve between $\mu_\mathrm{C}$ and $\mu_\mathrm{L}$
represents the electron flow from the left branch into the central
branch and correspondingly the area $A_\mathrm{R}$ between
$\mu_\mathrm{R}$ and $\mu_\mathrm{C}$ represents the electron flow
from the central branch to the right branch. Since the current
flowing into the junction has to equal the current
flowing out, and there is no net current flowing into or out of the
central branch, $A_\mathrm{L}$ and $A_\mathrm{R}$ have to be equal.
As a result the voltage at the center branch will always
be negative. Correspondingly it can be seen that for holes the voltage at the center branch will
always be positive. Following the same arguments we expect $V_\mathrm{C}$ as a function of
applied push-pull voltage $V_0$ to bend down for electrons and up
for holes. It should be noted that in this model the rectification
effect is a result of the change in conductance of the
device when the Fermi energy is changed. Therefore we expect the
rectification to be visible in graphene also for diffusive
transport.


\begin{figure}
  \begin{center}
    \includegraphics{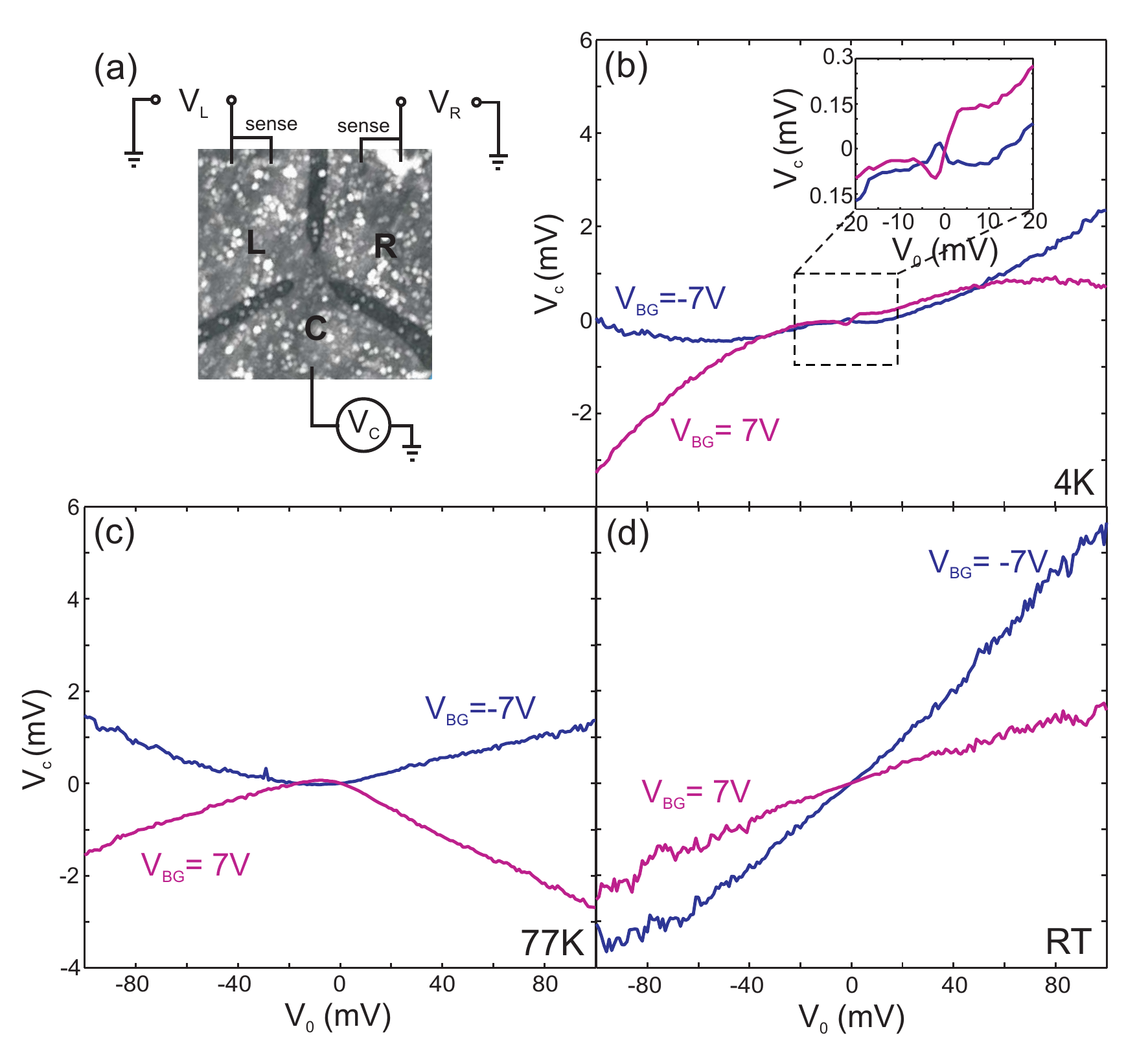}
    \caption{(a) The electrical measurement setup used in the non-linear regime. (b)-(d) $V_\mathrm{C}$ as a function of $V_0$
    for $V_{\mathrm{BG}}=+7$\,V and $V_{\mathrm{BG}}=-7$\,V at
     4\,K, 77\,K and room temperature respectively.}
    \label{fig2}
  \end{center}
\end{figure}

The measurement circuit used to investigate the voltage
rectification properties of the three-terminal junction in the
non-linear response regime is shown in Fig. \ref{fig2}(a). In order
to avoid contact resistances and to ensure that the bias
voltages are applied symmetrically to the device we have two Ohmic
contacts at each branch of the junction and use the sense function
of the DC voltage source to monitor the potential at one contact and
apply the potential at the second contact. The $V_0$ voltages plotted in
Figs.\,\ref{fig2}(b)-(d) are the measured voltages and not the applied
voltages, which are larger since they partially drop across the
contacts. With this setup we also eliminate possible non-linearities
of the contacts themselves.

Figs.\,\ref{fig2}(b)-(d) shows the measurement of $V_\mathrm{C}$ as a
function of $V_0$ for backgate voltages +7\,V (electrons) and
-7\,V (holes) at 4\,K, 77\,K and room temperature, respectively. As a
general trend we see that $V_\mathrm{C}$ as a function of $V_0$
always bends down for $V_{\mathrm{BG}}=+7$\,V and always bends up
for $V_{\mathrm{BG}}=-7$\,V. At 4\,K additional kinks can be seen in the curves
for $\left|{V_\mathrm{0}}\right| < 20$\,mV (see the inset in
Fig.\,\ref{fig2} (b)). This is due to universal conductance fluctuations
(UCFs) and has been theoretically predicted \cite{csontos2003}. The
linear part of $V_\mathrm{C}$ as a
function of $V_0$ is due the asymmetry of the device, which strongly depends on $V_{\mathrm{BG}}$.

\begin{figure}
  \begin{center}
    \includegraphics{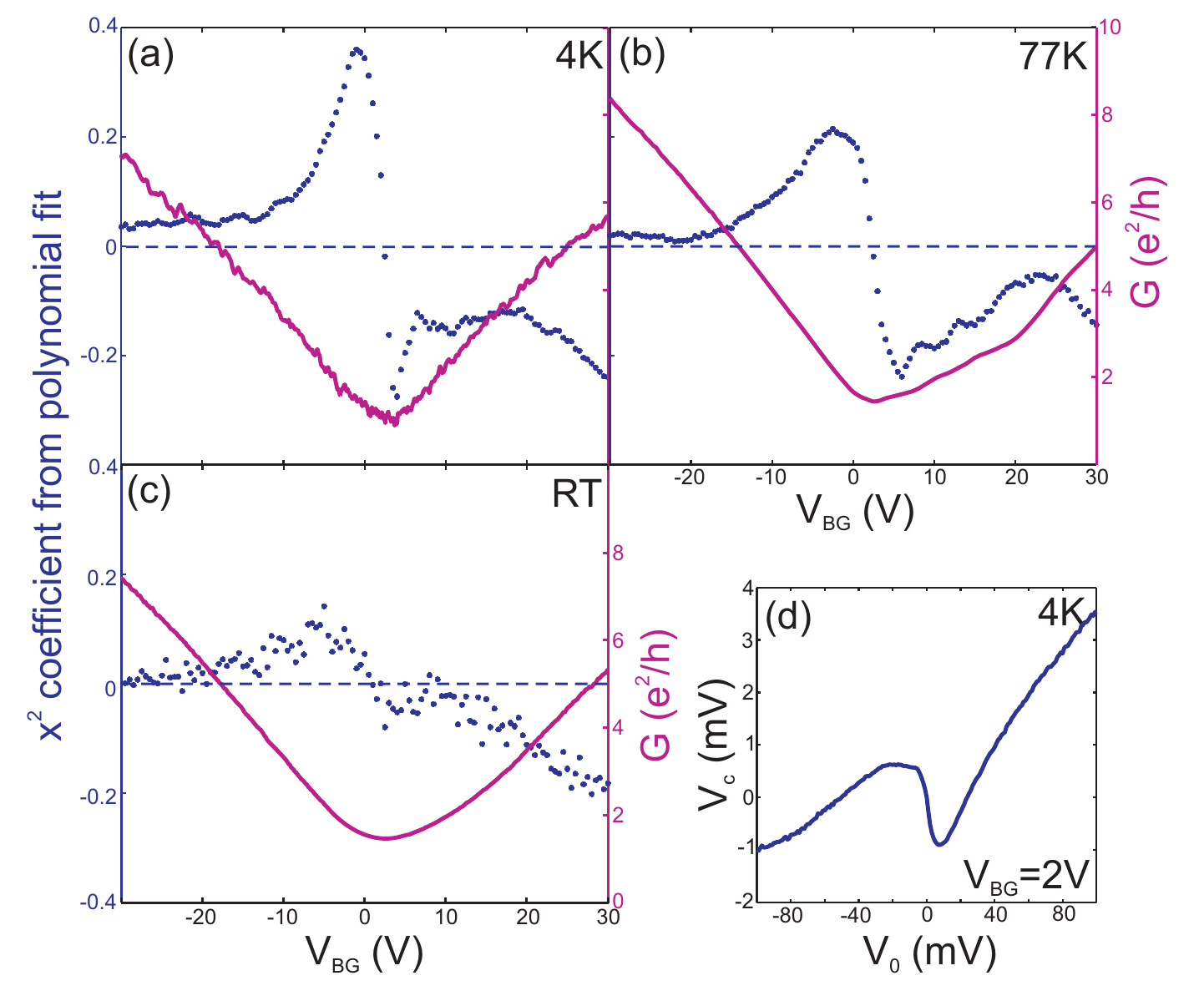}
    \caption{(a)-(c) The second-order coefficients obtained from fitting $V_\mathrm{C}$ vs
    $V_0$ with a third order polynomial (blue dots) and conductance from left to right (pink line) as a function of
     $V_{\mathrm{BG}}$ at 4\,K, 77\,K and room temperature respectively. (d) $V_\mathrm{c}$ vs $V_\mathrm{0}$ close to the Dirac point.}
    \label{fig3}
  \end{center}
\end{figure}

In the following we will have a closer look at the backgate
dependence of the rectification signal. From the simple model shown in
Fig.\,\ref{fig1}(c) it is expected that the $V_\mathrm{C}$ vs $V_0$
curves always bend down for electrons and bend up for holes as
confirmed in Figs.\,\ref{fig2}(b)-(d). However, it is also expected that the
magnitude of curvature will decrease with increasing Fermi energy, meaning that the further the device is tuned away from the
Dirac point the weaker the curvature is since the relative difference between the conductances
left and right of the junction is small compared to the overall conductance. In Figs.\,\ref{fig3}(a)-(c) the magnitude of curvature (blue dots) and the conductance
from the left lead to the right lead (pink curve) is plotted as a
function of $V_{\mathrm{BG}}$ at 4\,K, 77\,K and room temperature,
respectively. Here we have defined the magnitude of curvature as the
second-order coefficient obtained from fitting the $V_\mathrm{C}$ vs $V_0$
curve with a third order polynomial. It can be clearly seen that
for holes the curvature is positive and for electrons it is
negative, in agreement with the individual measurements shown in
Figs.\,\ref{fig2}(b)-(d). Measuring the device in any of the two other possible measurement configurations gives qualitatively the same results.
For holes it can be seen that the magnitude of
curvature decreases while going to higher charge carrier densities
as expected, however for electrons the curvature first decreases and then increases
with increasing charge carrier density. We
attribute this to inhomogenities in the sample leading to different
charge neutrality points at different locations in the sample and
therefore a mixture of electron and hole transport. This is
supported by the kinks seen in the $G$ vs $V_{\mathrm{BG}}$
curves, for instance the kink at approximately +20\,V backgate in
Fig.\,\ref{fig3}(b). In addition it can be seen that as a general trend the magnitude of curvature
decreases with increasing temperature, which is due to thermal smearing around the Fermi energy.
Close to the Dirac point the $V_\mathrm{C}$ vs $V_\mathrm{0}$ curves
cannot be fitted well presumably because of electron-hole
 puddles and strong disorder (see the example depicted in Fig.\,\ref{fig3}(d)) \cite{martin2008}.

For the measurements presented in this paper a voltage rectification of a few percent is obtained.
According to our model the amount of
rectification increases when the Fermi energy decreases and the difference
in relative conductance between the left and the right constriction increases. To improve the rectification properties of the
 nanojunctions one possibility is therefore to make the devices smaller in order to
 open a transport gap and then tune the Fermi energy to a point close to the gap.


To conclude, we have observed voltage rectification up to room temperature in a diffusive graphene
nanojunction when operated in the non-linear transport regime. Tuning the device from electron to hole transport
the sign of rectification is switched.


We thank T. Ihn for helpful discussions and the Swiss National Science
Foundation (SNF) for financial support.


\end{document}